\documentclass[%
aip,
amsmath,amssymb,
reprint,
final,%
]{revtex4-1}

\usepackage [english]{babel}
\usepackage [utf8]{inputenc}
\usepackage [T1]{fontenc}

\usepackage{newtxtext}
\usepackage[varvw]{newtxmath}

\usepackage{siunitx}
\usepackage{array}
\usepackage{booktabs}
\usepackage{graphicx}
\usepackage{hyperref}
\hypersetup{
	colorlinks=true,   
	linkcolor=blue,    
	citecolor=blue,    
	filecolor=magenta, 
	urlcolor=cyan      
}

\providecommand{\ifhighlighting}{\iffalse}  
\ifhighlighting
  \usepackage{xcolor}
  \newcommand{\hlstart}{\color{purple}}
  \newcommand{\hlend}{\color{black}}
\else
  \newcommand{\hlstart}{}
  \newcommand{\hlend}{}
\fi

\newcommand \vc[1]{ \mathbf{#1} }

\newcommand \uvec[1]{ \hat{\vc{#1}} }

\newcommand{\Tensor}[1]{
  \overset
    {\text{\tiny$\boldsymbol\leftrightarrow$}}
    {\mathbf{#1}}
}

\newcommand{\Id}{\openone}  

\newcommand \ket [1]{ | #1 \rangle }

\renewcommand \Re { \mathop{\mathrm{Re}} }
\renewcommand \Im { \mathop{\mathrm{Im}} }

\newcommand \Trace { \mathop{\mathrm{Tr}} }

\usepackage{mathtools}

\DeclarePairedDelimiterXPP \ConfAvg[1] {} \langle \rangle {_\mathrm{conf}} {#1}

\newcommand{\rme}{\mathrm{e}}
\newcommand{\rmi}{\mathrm{i}}

\newcommand{\rmg}{\mathrm{g}}
\newcommand{\rmL}{\mathrm{L}}

\newcommand{\JE}{J_{\rme}}
\newcommand{\JG}{J_{\rmg}}

\newcommand{\Du}{\vc{D}^{\dagger}}
\newcommand{\Dd}{\vc{D}}

\newcommand{\FLI}{\mathscr{F}}  
\newcommand{\Tcyc}{T_{\mathrm{cyc}}}  

\newcommand{\dif}[1]{\mathinner{\mathrm{d} #1}}

\newcommand{\Liou}{ \mathscr{L} }

\newcommand{\LiouG}{ \Liou_\gamma }
\newcommand{\LiouL}{ \Liou_{\mathrm{L}} }
\newcommand{\LiouInt}{ \Liou_{\mathrm{int}} }

\newcommand{\Eps}{ \boldsymbol{\varepsilon} }
\newcommand{\EpsL}{ \Eps_{\mathrm{L}} }

\DeclareMathOperator\erfc{erfc}

\newcommand{\PeakAmp}[3]{%
  A_{\uvec{#1},#2}(#3)%
}
\newcommand{\PeakAmpExplicit}[3]{%
  \Re\{S_{\uvec{#1},#2} (\ifx 1#3\else#3\fi \omega_0;#3)\}
}

\begin{document}
	
	
	\title[Pulse area dependence of multiple quantum coherence signals in dilute thermal gases]{Pulse area dependence of multiple quantum coherence signals in dilute thermal gases}

	\author{Benedikt Ames}%
	\affiliation{%
		Physikalisches Institut, Albert-Ludwigs-Universit\"at Freiburg, Hermann-Herder-Str.\ 3,
		D-79104 Freiburg, Germany 
	}%
	
		\author{Andreas Buchleitner}
	\affiliation{%
		Physikalisches Institut, Albert-Ludwigs-Universit\"at Freiburg, Hermann-Herder-Str.\ 3,
		D-79104 Freiburg, Germany 
	}%
\affiliation{EUCOR Centre for Quantum Science and Quantum Computing, Albert-Ludwigs-Universit\"at Freiburg, Hermann-Herder-Str.\ 3, D-79104 Freiburg, Germany}
	\date{\today}

	\author{Edoardo G. Carnio}
	\affiliation{%
		Physikalisches Institut, Albert-Ludwigs-Universit\"at Freiburg, Hermann-Herder-Str.\ 3,
		D-79104 Freiburg, Germany 
	}%
	\affiliation{EUCOR Centre for Quantum Science and Quantum Computing, Albert-Ludwigs-Universit\"at Freiburg, Hermann-Herder-Str.\ 3, D-79104 Freiburg, Germany}
	
	\author{Vyacheslav N. Shatokhin}
	\email{vyacheslav.shatokhin@physik.uni-freiburg.de}
	\affiliation{Physikalisches Institut, Albert-Ludwigs-Universit\"at Freiburg, Hermann-Herder-Str.\ 3,
		D-79104 Freiburg, Germany 
	}%
\affiliation{EUCOR Centre for Quantum Science and Quantum Computing, Albert-Ludwigs-Universit\"at Freiburg, Hermann-Herder-Str.\ 3, D-79104 Freiburg, Germany}
	
	\begin{abstract}
		In the general framework of open quantum systems, we assess the impact of the pulse area on single and double quantum coherence (1QC and 2QC) signals extracted from fluorescence emitted by dilute thermal gases. We show that 1QC and 2QC signals are periodic functions of the pulse area, with distinctive features which reflect the particles' interactions 
		via photon exchange, the polarizations of the laser pulses, and the observation direction. 
	\end{abstract}
	
	\maketitle
	
	
	
	
	
	


\section{Introduction}
\begin{figure}
  \includegraphics[width=0.45\textwidth]{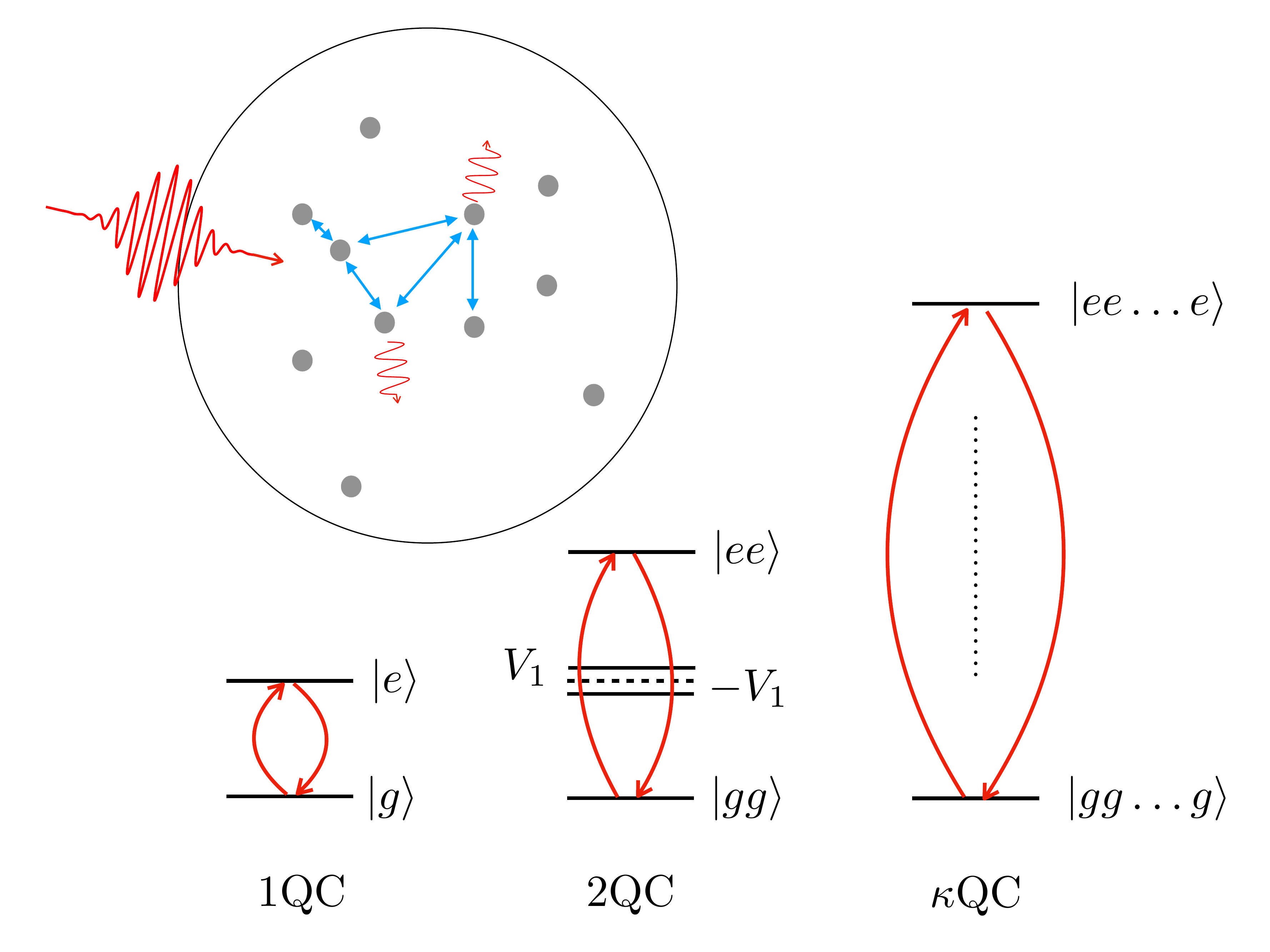}
  \caption{%
    An external field incident on a cloud of neutral atoms induces dispersive
    interactions between transition dipoles.
    So-called single quantum coherence (1QC) signals emitted by the atoms are
    probes of single-particle dipole moments (coherences between the ground and
    excited  states $\ket{g}$ and $\ket{e}$, respectively, of the
    individual atoms), while double quantum coherence (2QC) signals (in
    general, $\kappa$QC signals) are probes of a collective dipole moment of
    two (in general, $\kappa$) particles (coherence between the collective
    ground and excited state of a system of $2,\ldots, \kappa$ particles).
    $\kappa$QC ($\kappa>1$) signals are sensitive to shifts in the
    energy levels ($V_1$, $-V_1$ in the figure) induced by the
    dipole-dipole interactions (indicated by blue arrows), and
    therefore provide important information about collective
    excitations and interactions in many-body systems.
  }
  \label{fig:setup}
\end{figure}

If an ensemble of two-level atoms in their electronic ground states is driven by a resonant laser pulse, photoabsorption events will trigger \emph{dispersive}  interactions between transition dipoles of excited and unexcited particles\cite{stephen64,hutchison64}, which are of paramount fundamental and practical importance in atomic, molecular, and chemical physics. Through the particles' polarizability, these interactions modify the system's refractive index responsible for wave dispersion\cite{milonni_book}, hence their name. A common manifestation of dispersive interactions are collective shifts of the energy levels of a many-body system\cite{Friedberg1973101}. However, in a \emph{dilute thermal} gas these shifts are negligibly small compared to the Doppler broadening  introduced by thermal motion, rendering their sensing challenging. 
Nonetheless, ultrafast nonlinear optical spectroscopy\cite{mukamel_book} provides tools to observe subtle features of the coupling between neutral particles, via measurements of so-called multiple quantum coherence (MQC) signals, which are an indication of collective excitations in interacting systems\cite{Dai_2012,cundiff2013} (see Fig.~\ref{fig:setup}). In particular, it was shown that double-quantum coherence (2QC) signals appear only in the presence of level shifts induced by dipole-dipole interactions\cite{Gao:16}.

One of the most powerful methods to detect MQC signals in dilute thermal atomic ensembles is provided by non-linear two-dimensional electronic spectroscopy (2DES)\cite{Jonas2003,Bruder_2019A}. 
Although 2DES has proved to be a sensitive technique to probe dipolar interactions, it relies on measurements of the nonlinear response functions induced by a series of non-collinear, time-delayed fields\cite{Jonas2003,mukamel_book}. The interaction of the latter with a sample results in the emission of photon echo- or free-induction decay-type signals in directions satisfying phase-matching conditions\cite{Jonas2003}, which are different for distinct MQC signals\cite{mukamel_book}. However, fulfilling these conditions becomes increasingly more difficult at lower densities and for MQC signals of higher order.

This complication, as well as some experimental artifacts associated with 2DES\cite{Mueller:2019kd} can be overcome by fluorescence detection-based phase modulated spectroscopy \cite{tekavec2007}. In this method, one sends a series of collinear, phase-tagged pulses separated by one\cite{Bruder_2015,lukas_bruder15} or several\cite{tekavec2007,Yu:19a} time delays onto a sample, and collects the fluorescence signal emitted by excited atoms in the transverse direction. The detected signal encodes, in particular, different orders of MQC signals, which can be individually extracted via demodulation. With these techniques, successful measurements of $\kappa$QC signals in dilute thermal gases with $\kappa\leq 7$ were reported\cite{lukas_bruder15,Yu:19a,Bruder_2019}.
Since  MQC spectra in fluorescence-based measurements originate from excited state populations, rather than coherences of atoms, early results\cite{lukas_bruder15} initiated a debate\cite{Mukamel_2016,Li_2017,Bruder_2019,kuehn2020} whether the detected signals actually do certify dipole-dipole interactions in dilute, inhomogeneously broadened atomic clouds. 

Recently, we 
put forward a microscopic theory of MQC signals in dilute thermal gases which allowed us to resolve this controversy\cite{ames20}. Our theory includes several ingredients that have not been accounted for previously, such as the full form of the light-induced dipole-dipole interactions featuring both, \emph{near-field} (or \emph{electrostatic}) and \emph{far-field} (or \emph{radiative}) contributions, the vector character of the atomic dipole transitions as well as of the radiation field of the incoming laser pulses, and the unavoidable configuration (or disorder) average in a system of randomly located and mobile scatterers.   
We showed that interactions are actually crucial for the observations of $\kappa$QC signals, but they are mediated not by the hitherto considered\cite{Li_2017,Bruder_2019} electrostatic form of the dipole-dipole interactions, scaling as $\sim r^{-3}$, with $r$ the interatomic distance, but by the light-induced far-field dipolar interactions scaling as $\sim r^{-1}$. As distinct from the electrostatic interaction, which only generates collective level shifts, the far-field interaction mediates both, collective level shifts and collective decay processes, both of which are crucial for the emergence of $\kappa$QC signals. In the chemical physics literature, such far-field interactions via real photon exchange are known as cascading\cite{blank1999,Bennett:2014fp} or wave mixing processes\cite{gregoire2017}. Therefore, our results may also be beneficial for the understanding of multi-quantum coherence signals in multilevel isolated chromophores\cite{Mueller:2020bp}, where cascading events are a source of noise. 

A distinctive feature of our approach is its non-perturbative character with respect to the strength of the driving field. Previously\cite{ames20} we calculated MQC signals with parameters as in the experiment in Ref.\ \onlinecite{Bruder_2019}. In particular, the relatively weak laser pulses allowed us to treat the atom-laser interaction perturbatively, and to obtain good qualitative agreement with the experiment. Yet, it is well known that stronger pulses induce \emph{nonlinear atomic responses of higher order}\cite{mukamel_book}, which may lead to novel features of MQC signals. In the present work, we identify these features and ponder how they can be used to understand better the interplay between laser-atom and dipole-dipole interactions. In addition, we treat the atomic motion more accurately than in our earlier work\cite{ames20}, and thereby obtain not only qualitative, but also quantitative agreement of our simulations with the experimentally observed, Doppler broadened line shapes of the complex 1QC and 2QC spectra.

The paper is structured as follows: In the next section, we equip the reader with the relevant background information. Thereafter, in Sec.~\ref{sec:results}, we obtain inhomogeneously broadened 1QC and 2QC spectra in different observation directions and for different polarizations of the laser fields. Finally, we analyze the behavior of the spectral peaks as functions of the pulse area. Section \ref{sec:conclusions} concludes this work.

\section{Background}

We set out with a brief description of the experimental setup for the observation of MQC signals.  In Sec.~\ref{sec:meq}, we lay out our main theoretical tool, a master equation governing the dynamics of a multi-atom, dipole-dipole interacting system excited by laser fields. Section \ref{sec:solution} outlines  the main steps towards an analytical solution of the master equation. From this solution, the average fluorescence intensity is obtained upon configuration average, which is explained in Sec.~\ref{sec:dis}. Finally, in Sec.~\ref{sec:demod} we show how 1QC and 2QC spectra follow from the fluorescence signal upon demodulation.  

\subsection{Fluorescence detection-based phase-modulated spectroscopy}
\label{sec:spectroscopy}
In this spectroscopic approach, atoms are excited periodically by pairs of
collinear, time delayed, ultrashort Gaussian pulses. 
For
%
the $m$th cycle, beginning at time $\tau_m = m \Tcyc$ and with a
duration~$\Tcyc$, which is much longer than the natural lifetime of the atoms,
\hlend
the field in the frame rotating at the laser frequency $\omega_\rmL$ has the
time-dependent amplitude
\begin{equation}
  \vc{E}_\rmL (\vc{r},t)
	=
	\sum_{j=1}^2
	\EpsL^{(j)}
  \mathscr{E}_{\rmL}(t - t_j)
	\rme^{ \rmi(
    \vc{k}_{\rmL} \cdot \vc{r}
		+ \omega_{\rmL} t_j 
		+ \phi_j
		)}
	\, ,
	\label{laser}
\end{equation}
where
$
  \mathscr{E}_{\rmL}(t')
  =
  \mathscr{E}_0 \exp(-t'^2 / 2 \sigma^2)
$
is the envelope, assumed to be the same for both pulses, with amplitude
$\mathscr{E}_0$ and duration $\sigma$, and $\EpsL^{(j)}$ is the polarization of
pulse $j$ (in the following we will consider the cases
$\EpsL^{(1)} = \EpsL^{(2)} = \uvec{x}$ and
$\EpsL^{(1)} = \uvec{x}$, and
$\EpsL^{(2)} = \uvec{y}$).
Furthermore,
  both laser pulses are transmitted through acousto-optic modulators
  oscillating at slightly different frequencies~$w_j$.
  This continuous modulation imprints phase tags
  $\phi_j = w_j (\tau_m + t) \approx w_j \tau_m$ on the pulses,
  where the approximation is valid for femtosecond pulses and modulation frequencies
  $w_j \ll \omega_{\rmL}$\cite{Tekavec2006,Li_2017,beni_MSc}.
After the interaction of the atoms with the second pulse at $t_2$, the fluorescence
intensity along a direction
$\uvec{k} \perp \vc{k}_\rmL$
is integrated by a photodetector until 
the cycle ends.
Using a long pulse train with 
$M_{\rm cyc} \gg 1$ 
  cycles allows to sweep a broad range of
phase tags $\phi_j$ and interpulse delays $\tau = t_2 - t_1$, which is a
prerequisite for signal demodulation.  

The signal to be demodulated is the transient intensity
$I_{\uvec{k}}(\tau, t_\mathrm{fl}, \phi_1,\phi_2)$
integrated over the fluorescence time $t_\mathrm{fl}$\footnote{%
  To avoid confusion, we use a special notation $t_\mathrm{fl}$ for the
  fluorescence detection time, while we retain the notation $t$ for a general
  time variable. 
}
which is approximated by the integral:
\begin{equation}
	\bar{I}_{\uvec{k}}(\tau, \phi_1,\phi_2)
	=
	\int_{0}^{\infty} \dif{t_\mathrm{fl}}
	I_{\uvec{k}}(\tau,t_\mathrm{fl},\phi_1,\phi_2)
	\, .
	\label{intensity}
\end{equation}
This quantity remains congested with different harmonics of the modulation frequency, reflecting the interaction processes of atoms with the laser pulses as well as with one another. 
To 
select 
a specific modulation frequency component of the intensity, 
the recorded photocurrent $\bar{I}_{\uvec{k}}(\tau, \phi_1,\phi_2)$ 
is demodulated by multiplication with a reference
signal
$\rme^{-\rmi\kappa w_{21}\tau_m}$
($w_{21}=w_2-w_1$ the modulation frequency, and $\kappa$ the demodulation order -- not by accident the same label as the order of the MQC signal above)\cite{lukas_bruder15,Tekavec2006}.
The current 
at the thus identified modulation frequency is experimentally extracted by a lock-in narrowband filter, tantamount of integrating the spectral intensity, within a Lorentzian frequency
window of width $\sim \FLI$ centred around $\kappa w_{21}$, over the pulse train's duration $\tau_m$. Letting the filter width $\FLI\rightarrow 0$, the demodulated $\kappa$QC frequency spectrum 
is thus formally given by \cite{Tekavec2006,Li_2017}
\begin{eqnarray}
	S_{\uvec{k}}(\omega,\kappa)
	&=&
	\lim_{\FLI\to 0}\FLI \int_0^\infty \dif{\tau_m}\rme^{-(\FLI+\rmi\kappa w_{21}) \tau_m} \nonumber \\
	&\times& \frac{1}{\sqrt{2\pi}}\int_0^\infty \dif{\tau} \rme^{-\rmi\omega \tau}\bar{I}_{\uvec{k}}(\tau, \phi_1,\phi_2)	\,.
	\label{dem_signal}
\end{eqnarray}

\subsection{Physical system}
\label{sec:system}
We deal with a dilute thermal gas of alkali atoms at density
$n\sim \SI{e7}-\SI{e11}{\per\centi\meter\cubed}$
and temperature $\approx \SI{320}{K}$\cite{lukas_bruder15,Bruder_2019}.
Because of the thermal motion, a random spatial distribution of atoms in such a
cloud is time-dependent; the coordinate of atom $\alpha$ is given by
$
  \vc{r}_\alpha(t)
  =
  \vc{r}_{\alpha0} + \vc{v}_\alpha t
$,
where $\vc{r}_{\alpha0}$ and $\vc{v}_\alpha$ are,
respectively, its initial coordinate and its velocity.
We assume a homogeneous distribution of the initial coordinates
$\vc{r}_{\alpha0}$, while the Cartesian components of the velocities
$\vc{v}_\alpha$, drawn from the Maxwell-Boltzmann distribution for
the given temperature, are
typically~$\sim \SI[per-mode=symbol]{e2}{\meter\per\second}$. 
The atomic momentum at such velocities is several orders of magnitude larger
than the photon momentum, which allows us to neglect the photon recoil effect and
treat the atomic motion classically, i.e., ignore the coupling between the
atomic external and internal degrees of freedom.
Furthermore, because of the low atomic density $n$ satisfying the inequality
$n(\lambda/2\pi)^3\ll 1$ ($\lambda$~is the resonant optical wave length), the
atoms are typically in each other's far-field, such that inter-atomic
collisions can be neglected.
We treat the atoms as an open quantum system embedded into a common quantized
electromagnetic vacuum field.
The interaction of the atoms with the latter gives rise to effective dipole-dipole
interactions\cite{agarwal74}, upon tracing over the bath's degrees of freedom,
and the atomic internal dynamics are then described by a master equation (see
Sec.~\ref{sec:meq}).
Throughout this work, we unfold our formalism for a general system of $N$
atoms, but carry out all calculations for $N=2$.
A system of two atoms suffices to evaluate 1QC and 2QC
signals on which we focus in our present
contribution; in any case, our model\cite{ames20} includes all the essential
ingredients of the physics characterizing a laser-driven, dilute thermal atomic
ensemble 
representing an optically thin medium \cite{Lagendijk1996143}.
This is in contrast to the hypothesis \cite{Bruder_2019} that many-body
effects, such as delocalized excitations among numerous particles, mediated by
dipole-dipole interactions, are key to a quantitative understanding 1QC and 2QC
signals: In our present physical picture, delocalization requires higher-order
multiple scattering, which is improbable in optically thin systems
\cite{Lagendijk1996143}.

The $\kappa$QC spectra are calculated, via \eqref{intensity} and \eqref{dem_signal}, from the
transient fluorescent intensity
$I_{\uvec{k}}(\tau, t_\mathrm{fl}, \phi_1,\phi_2)$,
the quantum-mechanical expectation value of the
intensity operator, averaged over atomic configurations.
By definition\cite{glauber}, the time-dependent intensity is given by
\begin{equation}
	I_{\uvec{k}}(t)
	=
	\ConfAvg[\Big]{
		\bigl\langle
      \vc{E}^{(-)}_{\uvec{k}}(t) \cdot \vc{E}^{(+)}_{\uvec{k}}(t)
		\bigr\rangle
	}
	\label{int}
\end{equation}
where
$\langle\ldots\rangle = \Trace \{ \ldots \rho(0) \}$.
The initial density operator~$\rho(0)$ of the atoms-field system factorizes
into the ground state of the atoms and the vacuum state of the field, while the initial Heisenberg atomic and field operators are independent of each other. At later times $t>0$, the atom-light interaction couples the field and matter degrees of freedom, such that, in general, the field operators at any spatial point consist of a sum of the free field, the incident field, and the source field emitted by the atoms\cite{agarwal74}. In this work we study fluorescence in orthogonal directions to the wave vector of the incident pulses, such that the incident field cannot directly contribute to the detected fluorescence signal. Nor can the vacuum field of the electromagnetic bath. Thus, 
the positive/negative frequency part~$\vc{E}^{(+/-)}_{\uvec{k}}(t)$ of the
operator for the electric far-field solely stems from the scattered field, which 
can be expressed through the dipole operator of the source atoms $\alpha$\cite{agarwal74},
\begin{equation}
  \vc{E}^{(+)}_{\uvec{k}}(t)
	=
	f \sum_{\alpha=1}^N
	\Bigl\{
    \Dd_\alpha(t)
    -
    \uvec{k} \bigl[ \uvec{k}\cdot \Dd_\alpha(t) \bigr]
	\Bigr\}
  \rme^{-\rmi\vc{k}\cdot\vc{r}_\alpha(t)}
	\, ,
	\label{Eplus}
\end{equation}
where $f=\omega_0^2d/(4\pi\epsilon_0c^2R_d)$,
with $\omega_0$ the unperturbed atomic transition frequency (and $k_0$ the associated wave number), $d$ the transition
dipole matrix element, $c$ the speed of light, and $R_d$ the distance from the
center of mass of the atomic cloud to the detector. Moreover,
$\vc{k}$ is the wave
vector of the scattered light, and 
$\Du_\alpha$, ($\Dd_\alpha$) are  Heisenberg-picture  atomic raising (lowering) operators
which we choose to correspond to a $\JG=0\leftrightarrow \JE=1$ transition in the 
concrete calculations presented in Sec.~\ref{sec:results}.
Note, however, that all our subsequent expressions including dipolar operators are valid for
transitions incorporating {\em arbitrary} degeneracies of the ground and excited state
sublevels.

In writing~\eqref{Eplus}, we assumed that
$
  R_d\gg r_{\alpha\beta}(t)
  =
  | \vc{r}_\alpha(t) - \vc{r}_\beta(t) |
$
for any $\alpha$, $\beta$.
Finally, $\ConfAvg{\ldots}$ in \eqref{int} stands for the configuration
average.
It results from the random initial positions of the many pairs of atoms in
the cloud, and their thermal motion during the fluorescence detection time $t_\mathrm{fl}$.
We will treat the thermal velocities
in an effective manner, such that our results 
ultimately involve integrations over the Maxwell-Boltzmann distributions of the Doppler shifts of moving atoms, as well as 
over the length and orientation
of the vectors~$\vc{r}_{\alpha \beta}$ connecting pairs of atoms.
We denote the mean interatomic distance by~$\bar{r}$,
and assume an isotropic distribution of the atoms.

\subsection{Master equation}
\label{sec:meq}
With the aid of~\eqref{Eplus} the fluorescence intensity~\eqref{int} reads
\begin{equation}
  I_{\uvec{k}}(t)=\ConfAvg[\Big]{
    f^2\sum_{\alpha,\beta=1}^N
    \bigl\langle
        \Du_\alpha(t)\cdot (\Id-\uvec{k}\uvec{k})\cdot \Dd_\beta(t)
    \bigr\rangle
    \rme^{\rmi\vc{k}\cdot\vc{r}_{\alpha\beta}(t)}
  },
  \label{correlator}
\end{equation}
where $\Id$ is a $3\times 3$ unit matrix and $\uvec{k}\uvec{k}$ is a $3\times 3$ dyadic.
If the relative phase shift $k_0 v_\alpha r_{\alpha\beta}/c$ acquired by
moving atoms within the photon propagation time (given by
$r_{\alpha\beta}/c$) is small, i.e.,
$k_0v_\alpha r_{\alpha\beta}/c \ll 1$\cite{trippenbach92}, 
the time-dependent atomic dipole correlators entering Eq.~\eqref{correlator}
can be assessed via a Lehmberg-type master
equation for arbitrary $N$-atom operators\cite{PhysRevA.2.883} (or,
equivalently, for the $N$-atom density operator\cite{agarwal74}), under
the standard Born and Markov approximation\cite{breuer_book}.
This condition is well fulfilled in a thermal dilute gas with
$r_{\alpha\beta}\approx\bar{r}\sim \SI{10}{\micro\meter}$\footnote{%
  $\bar{r}\simeq \num{0.554} n^{-1/3}$,\cite{PhysRevA.49.146}
  which for a particle density
  $n\simeq \SI{e8}{\per\cm\cubed}$
  is about
  $\SI{10}{\micro\meter}$.
}
and 
$k_0v_\alpha\sim \SI{600}{\mega\hertz}$,
where
$
  k_0 v_\alpha r_{\alpha\beta}(t)/c
  \sim
  \num{2e-5}
$. 
We then obtain the same form of the dipole-dipole interaction as for immobile
atoms, where the usually fixed interatomic distance $r_{\alpha\beta}$ acquires a time dependence\cite{trippenbach92}.

An arbitrary $N$-atom operator
	$Q=\otimes_{\alpha=1}^NQ_\alpha$  obeys the quantum Heisenberg-Langevin equations \cite{cohen_tannoudji_API,PhysRevA.2.883,gremaud2006} 
\begin{equation}
	 \dot Q 
	=
(\Liou_0+\LiouL+\LiouG+\LiouInt) Q+F_Q(t),
	\label{HL}
\end{equation}
where $\Liou_0=\sum_{\alpha=1}^N\Liou_0^\alpha$, with $\Liou_0^\alpha$ the free-atom Liouvillian, 
$\LiouL=\sum_{\alpha=1}^N\LiouL^\alpha$, with $\LiouL^\alpha$ the laser-atom interaction Liouvillian,
$\LiouG=\sum_{\alpha=1}^N\LiouG^\alpha$, with $\LiouG^\alpha$ the relaxation Liouvillian describing the exponential decay of
atomic excited state populations and of coherences,
$\LiouInt=\sum_{\alpha\neq\beta=1}^N\Liou_{\alpha\beta}$, 
with $\Liou_{\alpha\beta}$ the dipolar atom-atom interaction Liouvillian, and $F_Q(t)$ the Langevin force operator proportional to an atomic operator $Q(t)$ multiplied by the photonic creation or annihilation operators at time $t=0$, to account for the vacuum fluctuations. 
 By bringing the operator $F_Q(t)$ to the normally-ordered form with respect to the initial field operators, we ensure that its partial trace over the field subsystem vanishes, $\Trace_F \{ F_Q(t) \rho(0) \}=0$. 
 \hlstart We further transform the resulting operator equation to the interaction picture with respect to the free-atoms Hamiltonian
 $H_0=\hbar\omega_0\sum_{\alpha=1}^N\Du_\alpha\cdot \Dd_\alpha$ (or, equivalently, the Liouvillian $\Liou_0$).\footnote{%
 	This transformation preserves the meaning of the quantum-mechanical expectation value $\langle\ldots\rangle$ and the fluorescence intensity \eqref{correlator}.}  Since this transformation 
 leaves the atomic correlators in Eq.~\eqref{correlator} invariant, and in order not to overburden our notation, we will retain the symbol $Q$ for the transformed $N$-atom operator. Furthermore, only the laser-atom interaction Liouvillian $\bar{\LiouL}=\rme^{\rmi H_0 t/\hbar}\LiouL \rme^{-\rmi H_0 t/\hbar}$ is modified in the interaction picture, while the Liouvillians $\LiouG$ and $\LiouInt$ preserve their form. 
 Hence, 
 an $N$-atom operator
$Q$ 
averaged over the initial field state obeys 
the equation: 
\begin{equation}
 \dot Q 
  =
  (\bar{\LiouL}+\LiouG+\LiouInt) Q.
  \label{meq}
\end{equation}
\hlend
The Liouvillians are given by\cite{shatokhin2005,shatokhin2006,PhysRevA.51.3128}
\hlstart
\begin{eqnarray}
  \bar{\LiouL^\alpha}
  Q_\alpha 
  &=&
  -\frac{\rmi d}{\hbar}
  \bigl[
    \Du_\alpha
    \cdot
    \vc{E}_{\rmL}(\vc{r}_{\alpha 0},t)\rme^{-\rmi \delta_\alpha t}
 \bigr.\nonumber\\
 &&\bigl.
   {} +
    \Dd_\alpha
    \cdot
    \vc{E}^*_{\rmL}(\vc{r}_{\alpha 0},t)\rme^{\rmi \delta_\alpha t}
    ,
    Q_\alpha
  \bigr]
  ,
  \label{L_a}
  \\
  \LiouG^\alpha
  Q_\alpha
  &=&
  \frac{\gamma}{2}
  \Bigl(
    \Du_\alpha
    \cdot
    \bigl[Q_\alpha,\Dd_\alpha\bigr]
    +
    \bigl[\Du_\alpha,Q_\alpha\bigr]
    \cdot
    \Dd_\alpha
  \Bigr)
  \nonumber
  ,
\end{eqnarray}
\hlend
with
\begin{equation}
	\delta_\alpha
  =
  \omega_{\rmL} - \omega_0 - \vc{k}_{\rmL} \cdot \vc{v}_\alpha
	\label{detuning}
\end{equation}
 the laser-atom detuning modified by the Doppler effect,
$\gamma$ the single-atom spontaneous decay rate, and
\begin{eqnarray}
  \Liou_{\alpha\beta}
  Q_{\alpha\beta}
  &=&
  \Du_\alpha
  \! \cdot \!
  \Tensor{T}(k_0 r_{\alpha\beta}(t),\uvec{n})
  \! \cdot \!
  \bigl[Q_{\alpha\beta}, \Dd_\beta\bigr]\nonumber\\
  &
  +
  &
  \bigl[\Du_\beta, Q_{\alpha\beta}\bigr]
  \! \cdot \!
  \Tensor{T}{}^*(k_0r_{\alpha\beta}(t),\uvec{n})
  \! \cdot \!
  \Dd_\alpha
  \, ,
  \label{L_ab}
\end{eqnarray}
with the light-induced dipole-dipole interaction tensor
$\Tensor{T}(k_0r_{\alpha\beta}(t),\uvec{n})$
for $\uvec{n}=\vc{r}_{\alpha\beta}/r_{\alpha\beta}$.
Upon substitution $\xi(t) =k_0r_{\alpha\beta}(t)$, the tensor
$\Tensor{T}(\xi(t), \uvec{n})$ reads
\begin{eqnarray}
  \Tensor{T} (\xi(t), \uvec{n})
 & = &
  \frac{3\gamma}{4} \rme^{-\rmi\xi(t)}
  \left[
    \frac{\rmi}{\xi(t)}
    \Bigl( \Id - \uvec{n}\uvec{n} \Bigr)\right.\nonumber\\  
   &&   \left.+
    \left(
      \frac{1}{\xi(t)^2}
      -
      \frac{\rmi}{\xi(t)^3}
    \right)
    \Bigl( \Id - 3 \uvec{n}\uvec{n} \Bigr)
  \right] .
  \label{tensor_T}
\end{eqnarray}
The real and imaginary parts of the tensor
$ \Tensor{T}=\Tensor{\Gamma}+\rmi \Tensor{\Omega}$
describe collective decay rates and collective level shifts\cite{agarwal74}.
In the far-field limit, $\xi \gg 1$, which we here consider, Eq.~\eqref{tensor_T} reduces to 
$
  \Tensor{T}(\xi,\uvec{n})
  \approx
  (3 \gamma / 4)
  g(\xi)
  [\Id - \uvec{n}\uvec{n}]
$,
with coupling parameter
$
  g(\xi)
  =
  \rmi\rme^{-\rmi\xi}/\xi
$,
$
  |g(\xi)| \ll 1
$.
In this regime, the collective decay rates and level shifts are, respectively,
\begin{equation}
  \Tensor{\Gamma}(\xi, \uvec{n})
  =
  \frac{3 \gamma}{4}
  \frac{\sin\xi}{\xi}
  (\Id - \uvec{n}\uvec{n})
  , \quad
  \Tensor{\Omega}(\xi, \uvec{n})
  =
  \frac{3 \gamma}{4}
  \frac{\cos\xi}{\xi}
  (\Id - \uvec{n}\uvec{n})
  ,
  \label{tensor_G_Om}
\end{equation}
and describe dipolar interactions via the exchange of transverse photons\cite{Andrews04}. 

\subsection{Analytical solution of the master equation}
\label{sec:solution}

\subsubsection{Separation of timescales}
\label{sec:timescales}
The possibility of solving Eq.~\eqref{meq} analytically is based on the
existence of several timescales characterising the system dynamics. 
The shortest timescale is set by the duration $\sigma \sim \SI{50}{\femto\second}$ of the laser pulses. 
The typical interpulse delay $\tau \sim \SI{10}{ps}$ determines the 
second timescale.
After the second pulse at time $t_2$, atomic fluorescence is monitored for times $t_\mathrm{fl}$, with
$0\leq t_\mathrm{fl}\leq \Tcyc - t_2 \sim 100\gamma^{-1}$,
until all atoms with the natural lifetime
$\sim \gamma^{-1}\approx \SI{30}{\nano\second}$\cite{steckRubidium87Line2015}
undergo a transition to their ground states, emitting photons. 
Thus, fluorescence detection establishes the long timescale, which typically
lasts for about five orders of magnitude longer than the interpulse delay --- for
a few microseconds. 
Furthermore, on that long timescale, the typical
thermal coherence time~$\tau_{\mathrm{th}}$\cite{PhysRevLett.97.013004}
of the order of $\SI{10}{\nano\second}$\footnote{%
  The thermal coherence time
  $\tau_{\mathrm{th}} = \lambda/v\approx\SI{10}{\nano\second}$
  can be obtained from the Doppler shift
  $k_0 v\approx\SI{560}{\mega\Hz}$ and the
  wavelength
  $\lambda=2\pi/k_0\approx\SI{790}{\nano\meter}$\cite{Bruder_2019}.
}
defines the regime of the atomic dynamics beyond which the Doppler effect is
not negligible.
In summary, the typical timescales satisfy the inequalities 
\begin{equation}
  \sigma
  \ll
  \tau
  \ll
  \underbrace{
    \tau_\mathrm{th}\lesssim \gamma^{-1}\ll \Tcyc
  }_{
    t_\mathrm{fl}
  }.
  \label{timescales}
\end{equation}
The existence of several timescales that differ by orders of magnitude allows us to find
approximate piecewise solutions of a simplified version of \eqref{meq}, which, on a given timescale, retains
only the relevant terms.
Thus, during the shortest timescale we keep only the Liouvillian $\LiouL$ in~\eqref{meq}.
In contrast, between the pulses and after the second pulse, that is, during the
intervals $\tau$ and $t_\mathrm{fl}$, we retain $\LiouG+\LiouInt$ and
ignore $\LiouL$.
We note that even though $\LiouG+\LiouInt$ is almost negligible during~$\tau$,
these terms introduce some dephasing, without which MQC spectra for
immobile atoms would represent delta peaks.
Finally, using the weakness of the dipolar coupling,
$|g(\xi)|\gamma\ll \gamma$, we solve~\eqref{meq} perturbatively with respect
to $\LiouInt$.

For for each subdomain of those piecewise solutions, we treat the atomic
motion according to how long the respective intervals are compared to the
thermal coherence time $\tau_\mathrm{th}$.
Since $\tau_\mathrm{th}\gg \sigma$, the atoms can be considered at fixed
positions during the atom-laser interaction.
For times exceeding $\tau_\mathrm{th}$, 
we instead perform a configuration average to account for the randomised atomic positions within  
the 
cloud (see Sec.~\ref{sec:dis}).


\subsubsection{Piecewise solutions}
\label{sec:piece}
We now present operator solutions 
of Eq.~\eqref{meq} 
pertinent to the different timescales,
	$\sigma$, $\tau$, and $t_\mathrm{fl}$ in~\eqref{timescales}, for some random
	atomic configuration, in order to combine them in the overall solution in Sec.~\ref{sec:combination}. 
The evaluation of quantum-mechanical expectation values, and of the average over atomic
positions and velocities of atoms, will follow in Sec.~\ref{sec:dis}. 

\paragraph{Interaction with laser pulses.}
\label{sec:laser}
During the shortest timescale~$\sigma$, it is sufficient to solve the
single-atom equation 
\hlstart
\begin{equation}
\dot{Q}_\alpha
  =
 \bar{\LiouL^\alpha} Q_\alpha,
  \label{eqLL}
\end{equation}
with $\bar{\LiouL^\alpha}$ \hlend given by~\eqref{L_a}. 
The action of the Liouvillian~\hlstart $\bar{\LiouL}$ \hlend acting on an $N$-atom operator~$Q$ is
then given by the tensor product of single-atom evolution operators.

Since the fast timescale associated with the laser pulses is much shorter than
all other timescales, we replace the Gaussian laser pulses with delta pulses
having the same pulse area.
Then the action of pulse~$j$ on an atomic operator~$Q_\alpha$ is reduced to an
instantaneous unitary transformation (kick) $R^\alpha_j$:
\begin{equation}
	Q_\alpha(t^+_j)
	=
	R^\alpha_j
	Q_\alpha(t_j^-)
	=
	\mathinner{
    \rme^{-\rmi\vartheta \mathscr{M}_j/2}
	}
	Q_\alpha(t_j^-)
	\mathinner{
    \rme^{\rmi\vartheta \mathscr{M}_j/2}
	}
	,
	\label{kick}
\end{equation}
where $Q_\alpha(t_j^-)$ ($Q_\alpha(t_j^+)$) is an atomic operator before (after) the kick, 
\begin{equation}
	\vartheta
	=
	\frac{2 d}{\hbar}	\int_{-\infty}^\infty \dif{t'}
  \mathscr{E}_{\rmL}(t'-t_j)
	,
	\label{pulse_area}
\end{equation}
is the pulse area, and $\mathscr{M}_j$ is an atomic operator,
\begin{equation}
  \mathscr{M}_j
  =
  S^\dagger \rme^{\rmi \varphi_j}
  +
  S \rme^{-\rmi \varphi_j },
	\label{redef_M_j}
\end{equation}
with
$S^\dagger = \Du_\alpha \cdot \EpsL$,
$S = \Dd_\alpha\cdot \EpsL^*$
the  components of atomic raising and lowering operators along the laser polarization, and 
the phase 
\begin{equation}
	\varphi_ j^{(\alpha)} =
  \vc{k}_{\rmL} \cdot \vc{r}_{\alpha 0}
	+ \omega_0 t_j
	+ w_j \tau_m 
	+
	\Delta_\alpha t_j,
	\label{varphij}
\end{equation}
 where
$
  \Delta_\alpha
  \equiv
  \vc{k}_\rmL\cdot\vc{v}_\alpha
  =
  k_\rmL v_z^{(\alpha)}
$
is the Doppler shift of atom~$\alpha$.

The transformation~\eqref{kick} is manifestly non-perturbative with respect to
the pulse area (laser field strength). 
Using the algebraic properties of the operators \hlstart $S^\dagger$ and $S$,  
which are very similar to those of the Pauli pseudo-spin operators $\sigma_+$ and $\sigma_-$ \cite{barnett_radmore}, \hlend 
we obtain the
following expansion of the superoperator~$R^\alpha_j$\cite{beni_MSc}:\footnote{The explicit expressions for $R^\alpha_j$  are rather cumbersome and can be found in Chap.~3.2 of Ref.~\onlinecite{beni_MSc}.}
\begin{equation}
	R^\alpha_j
	=
	\sum_{l=-2}^2
	\rme^{\rmi l \varphi_j^{(\alpha)}} R_j^{[l]}.
	\label{phase_R1}
\end{equation}
 with the $\varphi_j$-independent superoperators~$R_j^{[l]}$ encoding the
structure of the atomic dipole transition, the laser polarization, and the
pulse area\cite{beni_MSc}. 

Let us recall that each laser-atom interaction cycle begins with the atoms
prepared in their ground state $\ket{g}$.
In this case, the expansion of $R_1^\alpha$ in \eqref{phase_R1} takes a simpler form, which we label $R^\alpha_{1,g}$, and only describes the action of the \emph{first} pulse on
a \emph{ground-state} atom\cite{beni_MSc}:
\begin{equation}
	R^\alpha_{1,g}
  =
  \sum_{l=-1}^1
	\rme^{\rmi l \varphi_1^{(\alpha)}} R_{1,g}^{[l]}.
	\label{phase_R1_ground}	
\end{equation} 
 Since $l=-1,0,1$ in the above sum, the important consequence of~\eqref{phase_R1_ground} is that independent atoms cannot give rise to 2QC signals (see Sec.~\ref{sec:demod}), which require terms $l=\pm 2$ in the sum.

\paragraph{Radiative decay and dipolar interactions.}
\label{sec:spont}
\label{sec:single-atom-free-evolution}
As discussed above in Sec.~\ref{sec:timescales}, during the inter-pulse and
fluorescence harvesting intervals, respectively $\tau$ and
$T_\mathrm{cyc}-t_2$, we
solve the master equation
\begin{equation}
\dot Q 
  =
(\LiouG + \LiouInt) Q 
  \label{matr_L0_V}
\end{equation}
perturbatively with respect to~$\LiouInt$. 
Such solution can be most compactly represented with the aid of a Laplace transformations in the complex variable $z$, as 
\begin{equation}
\tilde{Q}(z)
	=
\sum_{n=0}^\infty
\mathscr{G}_\gamma(z)
	\Bigl[
    \LiouInt \mathscr{G}_\gamma(z)
	\Bigr]^n
	Q(t_0)
	\label{series_Qz}
\end{equation}
where
\begin{equation}
  \mathscr{G}_\gamma(z)
	=
	\int_0^\infty \dif{t}
  \rme^{-z t} \rme^{\LiouG t}
	=
	\frac{1}{z - \LiouG}
	\label{G_0}
\end{equation}
 is the resolvent superoperator, 
$
  \tilde{x}(z)
  =
  \int_{0}^\infty \dif{t'}
  \exp(-z t') x(t')
$,
and $Q(t_0)$ corresponds to the $N$-atom operator at the beginning of the interval under consideration, i.e.\ $t_0 = t_1$ for the evolution during the interpulse delay, and $t_0 = t_2$ when collecting the fluorescence after the second pulse.

In the following, we examine 1QC and 2QC signals for $N=2$ atoms that either
are not interacting at all or exchange a single photon.
These stem from 
single or double scattering contributions to the
fluorescence signal.
Formally, such contributions are described by the 
terms
$n=0,2$
in the series expansion~\eqref{series_Qz}.
By restricting ourselves to two atoms and double scattering, we exclude long
scattering paths, which is justified in an optically thin
medium\cite{jonckheere2000,shatokhin2006} as provided by a dilute thermal gas.
Furthermore, in such medium we can safely ignore recurrent
scattering processes\cite{Kupriyanov:2017oq}  in which a photon is bouncing
back and forth between the atoms\cite{PhysRevB.50.16729}.

\subsubsection{Overall solution}
\label{sec:combination}
When solving~\eqref{matr_L0_V} with the aid of the
Laplace transform, we introduce two distinct Laplace-transform variables $z_1$ and $z_2$ to describe, respectively, the evolution after the first pulse (over $\tau$) and after the second pulse (over $t_\mathrm{fl}$). In our stroboscopic model of the dynamics, we obtain the overall solution for a two-atom operator by combining expressions \eqref{phase_R1_ground} and \eqref{phase_R1}, for the first and second laser pulse,  with Laplace transforms of the evolutions during the interpulse delay and detection period \eqref{series_Qz}. The resulting two-dimensional Laplace image reads, at order $m$ in the dipolar coupling,
\begin{widetext}
  \begin{equation}
        \tilde{Q}^{[m]}(z_1,z_2,\varphi_1^{(1)},\varphi_1^{(2)},\varphi_2^{(1)},\varphi_2^{(2)})
    =
        \sum_{n=0}^m
        \mathscr{G}_\gamma(z_2)
        \Bigl[
          \LiouInt
          \mathscr{G}_\gamma(z_2)
        \Bigr]^{m - n}
        \mathinner{
          R_2^{\rmL}
        }
        \mathscr{G}_\gamma(z_1)
        \Bigl[
          \LiouInt
          \mathscr{G}_\gamma(z_1)
        \Bigr]^n
        \mathinner{
          R_{1,g}^{\rmL}
        }
        Q(0)
    ,
      \label{Q2}
  \end{equation}
\end{widetext}
 where
$R^\rmL_2=\otimes_{\alpha=1}^2 R^\alpha_2$,
$R^\rmL_{1,g}=\otimes_{\alpha=1}^2 R^\alpha_{1,g}$,
and $m=0,1,2$.
Note that
the phase tags~$\varphi_1^{(1,2)}$, $\varphi_2^{(1,2)}$  in the argument of the LHS of Eq.~\eqref{Q2}
are absorbed in $R_{1,g}^{\rmL}$ and $R_2^{\rmL}$ via~\eqref{phase_R1_ground} and~\eqref{phase_R1}, respectively.
Evaluation of the relevant atomic dipole correlators entering~\eqref{correlator}
according to Eq.~\eqref{Q2}, followed by the configuration average
$\ConfAvg{\ldots}$, yields the two-dimensional Laplace image 
$\tilde{I}_{\uvec{k}}(z_1,z_2,\varphi_1^{(1)},\varphi_1^{(2)},\varphi_2^{(1)},\varphi_2^{(2)})$ 
of the transient fluorescence intensity. 

\subsection{Configuration average}
\label{sec:dis}
The configuration average is crucial to capture essential properties of a thermal
atomic cloud, and to obtain reasonable agreement with experiment. 
In a thermal gas, atoms have random, time-dependent positions.
Thus, the averaging accounts for the intrinsic randomness of the atomic
positions in a disordered medium, as well as for their changes during
fluorescence harvesting (section~\ref{sec:timescales}).

The configuration average $\ConfAvg{\ldots}$ in  the RHS of ~\eqref{correlator}
allows us to draw important conclusions about the structure of the function
$\tilde{I}_{\uvec{k}}(z_1,z_2,\varphi_1^{(1)},\varphi_1^{(2)},\varphi_2^{(1)},\varphi_2^{(2)})$. %
  Our general strategy is to identify and remove all terms which are
  quickly oscillating as a function of the atomic coordinates.
  The configuration average of the simplified, slowly varying expression can
  be obtained by substituting the mean interatomic distance $r_{12} \to \bar{r}$
  and analytically computing the isotropic average over the
  orientation~$\uvec{n}$.

  First of all,
the
 fluorescence signal 
cannot 
retain
 spatial interference  contributions  stemming  from two-atom correlators in Eq.~\eqref{correlator}. 
  These
  are multiplied by the phase factors
  $\exp(\rmi {\bf k}\cdot{\bf r}_{\alpha\beta}(t))$ ($\alpha\neq \beta$),
  which disappear because of the homogeneous distribution of interatomic
  distances $r_{\alpha\beta}(t)$.
  Therefore, %
    to evaluate the fluorescence intensity~\eqref{correlator},
    we remove all terms with $\alpha \neq \beta$.

%
  Second, by virtue of Eqs.~\eqref{phase_R1}, \eqref{phase_R1_ground}, and
  \eqref{Q2}, 	each of the terms contributing to the single-atom correlators
  in Eq.~\eqref{correlator} has the form%
    \begin{equation}
      \ConfAvg[\Big]{
	C(z_1, z_2)
	\prod_{\alpha = 1, 2}
	\exp \Bigl(
	  \rmi 
	    l_\alpha \varphi_2^{(\alpha)}
	    + \rmi m_\alpha \varphi_1^{(\alpha)}
	\Bigr)
      },
      \label{corr_av}
    \end{equation}
    where
    $C(z_1, z_2)$
    is a function
    which is independent of the phases~$\varphi_j^{(\alpha)}$,
    and the integer coefficients take values
    $l_\alpha = 0,\pm1,\pm 2$
    and
    $m_\alpha = 0,\pm 1$.
  Since, by~\eqref{varphij}, the phases $\varphi_j^{(\alpha)}$ are proportional
  to the initial coordinates $\vc{r}_{\alpha 0}$,
    a correlation function of the form~\eqref{corr_av} can only survive
    the disorder average if the
    multiplicities $l_\alpha$ and $m_\alpha$
    of these phase factors
    satisfy the condition
    $l_\alpha + m_\alpha = 0$.
    Therefore, \eqref{corr_av} simplifies to
    \begin{equation}
      \ConfAvg[\Big]{
	C(z_1, z_2)
	\exp \Bigl(
	  \rmi l_1 \varphi_{21}^{(1)}
	  + \rmi l_2 \varphi_{21}^{(2)}
	\Bigr)
      },
    \end{equation}
    with $l_1,l_2=0,\pm 1$, which only depends on the phase differences
  \begin{equation}
    \varphi^{(\alpha)}_{21}
    =
    \varphi_{2}^{(\alpha)}-\varphi_{1}^{(\alpha)}
    =
    \omega_0\tau+w_{21}\tau_m+\Delta_\alpha \tau
    .
  \end{equation}
    Henceforth, we denote the Laplace image of the fluorescence signal by
    $\tilde{I}_{\uvec{k}}(z_1,z_2,\varphi^{(1)}_{21},\varphi^{(2)}_{21})$.
   %

Third,  the function
$\tilde{I}_{\uvec{k}}(z_1,z_2,\varphi^{(1)}_{21},\varphi^{(2)}_{21})$
 cannot  depend on terms linear in
$\LiouInt\propto g(\xi)=\rmi\rme^{-\rmi\xi}/\xi$,
since $\xi=k_0r_{\alpha\beta}$. 
However, terms which are independent of, or quadratic in, $\LiouInt$ yield
single and double scattering contributions proportional to  $|g(\bar{\xi})|^0=1$ and
$|g(\bar{\xi})|^2=1/\bar{\xi}^2$, respectively, where $\bar{\xi}=k_0\bar{r}$, with $\bar{r}$
the mean interatomic distance.
Note that, although the function
$|g(\xi)|^2=|\vc{r}_{\alpha0}-\vc{r}_{\beta 0}+(\vc{v}_{\alpha}-\vc{v}_\beta)t|^{-2}$
depends on the atomic velocities, we assume that the atoms remain in the
far-field of each other, such that  $|g(\xi)|^2$ varies smoothly with $\xi$,
justifying the replacement $|g(\xi)|^2\rightarrow |g(\bar{\xi})|^2$ in the
double scattering contribution. 
 Finally, 
among terms that are quadratic in $\LiouInt$, only particular
products of the matrix elements of $\Tensor{T}(\bar{\xi}, \uvec{n})$ survive
the averaging over the isotropic distribution of
the vector~$\uvec{n}$.\cite{ames2021}. 

Decomposing
$\tilde{I}_{\uvec{k}}(z_1,z_2,\varphi^{(1)}_{21},\varphi^{(2)}_{21})$
into single and double scattering contributions (furnished with superscripts
`[0]' and `[2]', respectively, from the contributions of order $[m]$ in \eqref{Q2}, to each correlator entering \eqref{correlator}), we obtain
\begin{equation}
  \tilde{I}_{\uvec{k}}(z_1,z_2,\varphi^{(1)}_{21},\varphi^{(2)}_{21})
  =
  \tilde{I}^{[0]}_{\uvec{k}}(z_1,z_2,\varphi^{(1)}_{21})
  +
  \tilde{I}^{[2]}_{\uvec{k}}(z_1,z_2,\varphi^{(1)}_{21},\varphi^{(2)}_{21}),	
  \label{I_I0_I2}
\end{equation}
with
\begin{widetext}
  \begin{equation}
    \tilde{I}^{[0]}_{\uvec{k}}(z_1,z_2,\varphi^{(1)}_{21})
    = 
    \sum_{l=-1}^1 \tilde{I}^{[0]}_{l;\uvec{k}}(z_1,z_2)
    \mathinner{
      \rme^{\rmi \varphi^{(1)}_{21} l}
    }
    ,\quad
    \tilde{I}^{[2]}_{\uvec{k}}(z_1,z_2,\varphi^{(1)}_{21},\varphi^{(2)}_{21})
    =
    \sum_{l_1=-1}^1
    \sum_{l_2=-1}^1
    \tilde{I}^{[2]}_{l_1,l_2;\uvec{k}}(z_1,z_2)
    \mathinner{
      \rme^{\rmi \varphi^{(1)}_{21} l_1 + \rmi \varphi^{(2)}_{21} l_2},
    }
    \label{I^2_expansion}
  \end{equation}
\end{widetext}
 where we used the invariance of \eqref{I_I0_I2} under permutations of the atoms, and considered the fluorescence signal emitted solely by atom 1 (note the phase $\varphi^{(1)}_{21}$ of $\tilde{I}^{[0]}_{\uvec{k}}(z_1,z_2,\varphi^{(1)}_{21})$ in \eqref{I^2_expansion}). 
Thus, the single and double scattering intensities $\tilde{I}^{[0]}_{\uvec{k}}(z_1,z_2,\varphi^{(1)}_{21})$ and $\tilde{I}^{[2]}_{\uvec{k}}(z_1,z_2,\varphi^{(1)}_{21},\varphi^{(2)}_{21})$ are defined up to a common prefactor of 2. 
The phase factors in~\eqref{I^2_expansion} (and those alone) depend on the
Doppler shifts $\Delta_\alpha$ as 
\begin{equation}
  \rme^{\rmi \varphi^{(\alpha)}_{21} l}
  =
  \rme^{\rmi (\omega_0\tau + w_{21}\tau_m + \Delta_\alpha\tau)l}.
  \label{phase_factor}	
\end{equation}	 
It remains to
integrate Eq.~\eqref{I^2_expansion} over the one-dimensional Maxwell-Boltzmann distributions,
\begin{equation}
  p(\Delta_\alpha)
  =
  \frac{1}{\bar{\Delta}\sqrt{2\pi}}
  \exp\left(-\frac{\Delta_\alpha^2}{2\bar{\Delta}^2}\right),
  \label{MBdistr}
\end{equation}	
where $\bar{\Delta}=(k_\rmL^2k_BT/M)^{1/2}$ is the root-mean-square (r.m.s.)
Doppler shift, with $T$ the temperature and $M$ the atomic mass. To that end, 
we evaluate the MQC spectra for
fixed, random Doppler shifts and, then, average the spectra over the
shifts' distribution \eqref{MBdistr}. 
This allows us to show, in the next subsection, how the homogeneous profiles of MQC spectra for
immobile atoms transform into the inhomogeneously broadened ones observed in
thermal vapors\cite{Bruder_phd}.

\subsection{Signal demodulation}
\label{sec:demod}
The phase modulation explained in Sec.~\ref{sec:spectroscopy} affects the expression $\tilde{I}_{\uvec{k}}(z_1,z_2,\varphi^{(1)}_{21},\varphi^{(2)}_{21})$ as given by \eqref{I_I0_I2}, \eqref{I^2_expansion} through the phase factors \eqref{phase_factor}. Therefore, the fluorescence signal is oscillating at harmonics of the frequency $w_{21}$.
The demodulation procedure allows one to extract specific frequency components
from the full signal.  
The corresponding spectra, generally referred to as multiple quantum coherences
(MQC), are given by Eqs.~\eqref{intensity} and~\eqref{dem_signal}.
Merging these equations into one formula, replacing the  phase tags~$\phi_1$
and~$\phi_2$ with the phase differences~$\varphi^{(\alpha)}_{21}$, and averaging
over the Doppler shifts' distribution \eqref{MBdistr}, we obtain 
\begin{eqnarray}
	S_{\uvec{k}}(\omega; \kappa)
	&=&
	\lim_{\FLI \to 0}
  \frac{\FLI}{\sqrt{2\pi}}
  \int_0^\infty \dif{\tau_m}
	\rme^{-(\FLI+\rmi \kappa w_{21})\tau_m}
	\int_0^\infty \dif{\tau}
	\rme^{-\rmi \omega \tau}
	\nonumber \\
	&
	\times
	&
  \int_{0}^\infty \dif{t_\mathrm{fl}}
	\int_{-\infty}^{\infty}\dif{\Delta_1} p(\Delta_1)
	\int_{-\infty}^{\infty}\dif{\Delta_2}p(\Delta_2)
  \nonumber\\
	&
	\times
	&
  I_{\uvec{k}}(\tau, t_\mathrm{fl}, \varphi^{(1)}_{21}(\Delta_1),\varphi^{(2)}_{21}(\Delta_2)), 
	\label{dem_signal_explicit1}
\end{eqnarray}
where $\kappa=1, 2$ in this work, and
$I_{\uvec{k}}(\tau, t_\mathrm{fl}, \varphi^{(1)}_{21},\varphi^{(2)}_{21})$ 
is the two-dimensional inverse Laplace transform of
$\tilde{I}_{\uvec{k}}(z_1,z_2,\varphi^{(1)}_{21},\varphi^{(2)}_{21})$;
thus, it has exactly the same dependence on the phases
$\varphi^{(1)}_{21}$, $\varphi^{(2)}_{21}$. 
By virtue of \eqref{phase_factor},  this filtering procedure selects terms of the intensity
modulated as
$\rme^{\rmi (\omega_0 +\Delta_1)\tau}$ for $\kappa=1$, and those 
modulated as
$\rme^{\rmi (2\omega_0 +\Delta_1+\Delta_2)\tau}$, for $\kappa=2$.

We first assess~\eqref{dem_signal_explicit1} by changing the integration order,
to take the integrals over $\Delta_1$ and $\Delta_2$ at the very end.
This allows us to express $S_{\uvec{k}}(\omega;\kappa)$ as the average over MQC
spectra of atoms with fixed velocities.
The latter spectra follow directly from the Laplace image
$\tilde{I}_{\uvec{k}}(z_1,z_2,\varphi^{(1)}_{21},\varphi^{(2)}_{21})$,
upon the replacements
$z_1\rightarrow \rmi(\omega-\omega_0-\Delta_1)$
and
$z_1\rightarrow \rmi(\omega-2\omega_0-\Delta_1-\Delta_2)$,
for $\kappa=1$ and $\kappa=2$, respectively, and $z_2\rightarrow 0$ because of the infinite upper integration limit in \eqref{intensity}.

\begin{figure*}
	\includegraphics[width=\textwidth]{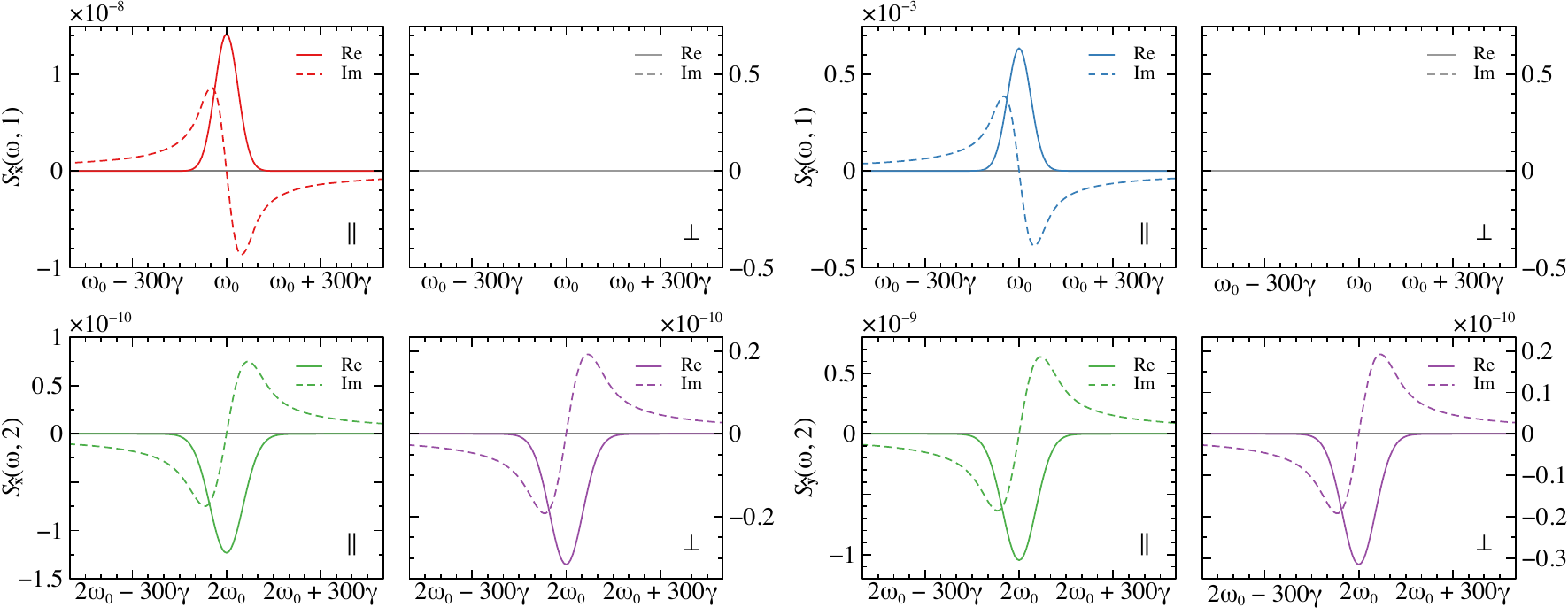}
	\caption{
		Real ($\Re$, solid lines) and imaginary ($\Im$, dashed lines) parts of the
		MQC spectra		
		$S_{\uvec{k}} (\omega;\kappa)$
		[in units of the spectral density integrated over the fluorescence detection time,
    $f^2/\gamma^2$; see~\eqref{correlator} and~\eqref{dem_signal_explicit1}] of $N=2$ atoms,
		for detection direction $\uvec{k}=\uvec{x}$ (left), 
		$\uvec{k}=\uvec{y}$ (right),
		and for quantum coherences of order $\kappa=1$ (top), $\kappa=2$ (bottom).
		Symbols $\parallel$ and $\perp$ indicate
		parallel ($\uvec{x}$-$\uvec{x}$)
		and perpendicular ($\uvec{x}$-$\uvec{y}$)
		pump-probe polarizations, respectively.
    All plots were obtained for temperature $T=\SI{320}{K}$, average
    interatomic distance $\bar{r}\approx \SI{10}{\micro\meter}$ (which
    corresponds to the particle density $\SI{e8}{\per\cm\cubed}$),
		laser pulse areas $\vartheta=\num{0.14}\pi$ and durations
		$\sigma= \SI{21}{fs}$ (tuned to resonance with the D2-line
    of $^{87}$Rb atoms\cite{steckRubidium87Line2015}), atomic mass
		$M=\SI{1.443e-25}{kg}$, transition wavelength $\lambda_0=\SI{790}{nm}$,
    and the spontaneous decay rate
    $\gamma \approx 2\pi \times \SI{6,067}{MHz}$, faithfully matching the
    experimental conditions\cite{Bruder_2019}.
    This choice of parameters results in a mean scaled interatomic distance
    $\bar{\xi}=k_0\bar{r}\approx 80$, with $k_0$ the atomic transition's wave number,
		and a r.m.s.\ Doppler shift $\bar{\Delta}=2\pi \times\SI{0,221}{GHz}$.
	}
	\label{fig:spectra}
\end{figure*}

The selection of terms in~\eqref{dem_signal_explicit1} with a specific
modulated phase due to~\eqref{I^2_expansion}, \eqref{phase_factor} implies that
only certain sum indices in~\eqref{I^2_expansion} can contribute to $\kappa$QC
signals:
The condition $l=\kappa$ and $l_1+l_2=\kappa$ must be fulfilled for the
single and double scattering contributions, respectively. 
Therefore, using Eqs.~\eqref{I_I0_I2}--\eqref{dem_signal_explicit1}, we obtain
that the 1QC and 2QC signals are given by the following convolutions of Lorentzian
(for a given velocity/Doppler shift)
and Gaussian distributions~\eqref{MBdistr},
\begin{eqnarray}
	S_{\uvec{k}}(\omega; 1)
  &=&
  \int_{-\infty}^{\infty}\dif{\Delta_1}p(\Delta_1)
  \left[
    \tilde{I}^{[0]}_{1;\uvec{k}}\Bigl(\rmi (\omega-\omega_0-\Delta_1),0\Bigr)
  \right. \nonumber\\
	&
  +
	&
	\left.
    \tilde{I}^{[2]}_{1,0;\uvec{k}}\Bigl(\rmi (\omega-\omega_0-\Delta_1),0\Bigr)
  \right]
  , \label{S1}
  \\
	S_{\uvec{k}}(\omega; 2)
  &=&
	\int_{-\infty}^{\infty}\dif{\Delta_1}p(\Delta_1)
	\int_{-\infty}^{\infty}\dif{\Delta_2}p(\Delta_2)\nonumber\\
	&
	\times
	&
	\tilde{I}^{[2]}_{1,1;\uvec{k}}\Bigl(\rmi (\omega-2\omega_0-\Delta_1-\Delta_2),0\Bigr)
  . \label{S2}
\end{eqnarray}
The above expressions yield complex $\kappa$QC spectra with resonances at
$\omega=\kappa \omega_0$; the real, absorptive parts of the spectral line
shapes are also known as Voigt profiles\cite{mukamel_book}. 
If we take the limit $T\to 0$ in~\eqref{MBdistr},
the distribution $p(x)\to \delta(x)$, and the expressions~\eqref{S1} and~\eqref{S2}
for $\kappa$QC spectra reduce to the complex Lorentzians reported elsewhere\cite{ames20}. 
Equation \eqref{S1} indicates that 1QC spectra can emerge from either independent or interacting atoms, whereas equation \eqref{S2} \emph{only} contains double scattering contributions, such that 2QC spectra \emph{do} require interactions via photon exchange.

Next, based on the above results, we examine 1QC and 2QC spectra for different
pump-probe polarizations, observation directions, and pulse areas.

\section{Results and discussion}
\label{sec:results}
\subsection{Inhomogeneously broadened 1QC and 2QC spectra}
\label{sec:spectra}
Using symbolic computation software, we analytically assess the single
and double quantum coherence spectra $S_{\uvec{k}}(\omega;1)$ and
$S_{\uvec{k}}(\omega;2)$ of $N=2$ atoms,
as given by Eqs.~\eqref{S1} and~\eqref{S2}, for arbitrary pulse areas $\vartheta$,
for detection directions $\uvec{k}=\uvec{x}, \uvec{y}$,
and for parallel ($\uvec{x}$-$\uvec{x}$) and perpendicular ($\uvec{x}$-$\uvec{y}$)
pump-probe polarization channels (in brief, $\parallel$ and $\perp$ channels).
First, however, we consider parameter values close to the experimental ones \cite{Bruder_2019}, which, in particular, correspond to a fixed pulse area $\vartheta = 0.14\pi$. The spectra that result in this case are shown in Fig.~\ref{fig:spectra}.

\begin{figure}
	\includegraphics[width=\columnwidth]{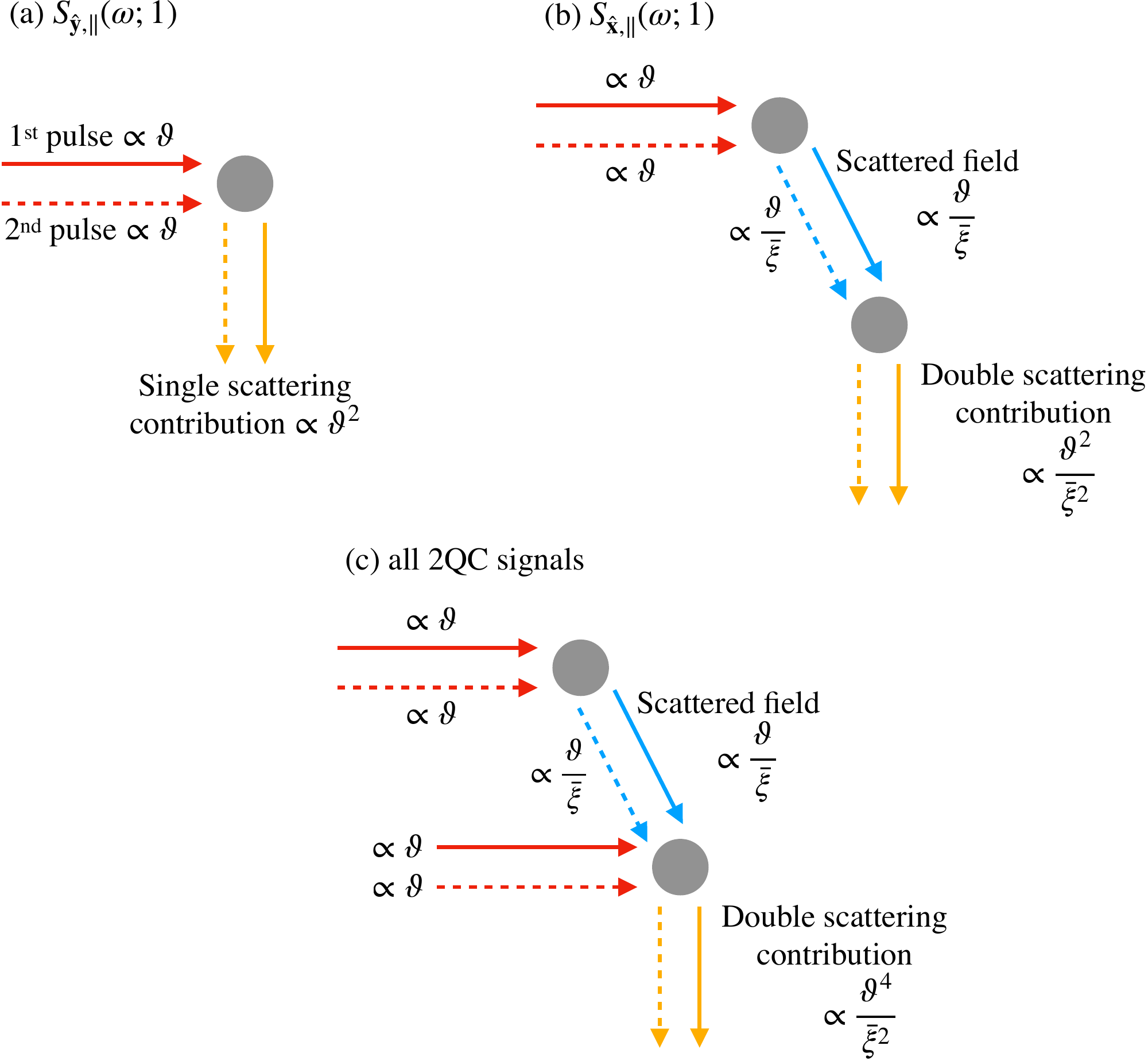}
	\caption{
	Scaling of 1QC signals (a) $S_{\uvec{y},\parallel}(\omega;1)$, (b) $S_{\uvec{x},\parallel}(\omega;1)$, as well as all (c) 2QC signals, with the pulse area $\vartheta$ and the mean scaled interatomic distance $\bar{\xi}$, in the weak,
	$\vartheta \ll 1$, and 
	far-field, $\bar{\xi}\gg 1$, limit. Signal (a) originates from single scattering, 
	signals (b,c) from double scattering. 		
		Horizontal red lines indicate the incoming laser pulses, vertical yellow lines the emitted fluorescence, oblique blue lines the scattered fields via dipole-dipole interactions. 
		Solid and dashed lines distinguish 
		a single field amplitude from its complex conjugate.}
	\label{fig:scattering}
\end{figure}


Whenever they do not vanish identically, the complex 1QC and 2QC spectra
feature, respectively, absorptive and dispersive resonances of the real and
imaginary parts. 
We recall that, by \eqref{S1}, 1QC signals can result from single scattering (one atom interacts with two laser fields, Fig.~\ref{fig:scattering} (a)) as well as from double scattering (one atom driven by two fields scattered by the other atom, Fig.~\ref{fig:scattering} (b)).
In the case of $S_{\uvec{y},\parallel}(\omega;1)$, single scattering dominates the signal, since both incoming pulses are polarized along the $\uvec{x}$-axis, while double scattering only yields a small correction. 
In contrast, non-interacting atoms driven by $\uvec{x}$-polarized fields cannot emit in the $\uvec{x}$ direction. The signal can then only emerge from the $\uvec{y}$ and $\uvec{z}$ components of the transition dipoles, which can be excited via double scattering, wherefrom the $S_{\uvec{x},\parallel}(\omega;1)$ signal stems.
In the weak-field limit, 
the 1QC spectrum $S_{\uvec{y},\parallel}(\omega;1)$ scales as $\vartheta^2$, whereas $S_{\uvec{x},\parallel}(\omega;1)$ scales as
$\vartheta^2/\bar{\xi}^2$, so that both are proportional to the linear susceptibility\cite{tekavec2007}.  
By \eqref{S2}, the 2QC spectra always originate from double scattering, where, in this context, an atom is
subject to two laser fields and two fields scattered by the other atom (Fig.~\ref{fig:scattering} (c)). In the
weak-field limit, the 2QC spectra scale with the pulse area as $\vartheta^4/\bar{\xi}^2$,
and are therefore proportional to the third-order non-linear susceptibility\cite{mukamel_book,boyd_book}.
Since the signs of the first- and third-order susceptibilities are opposite\cite{boyd_book}, so
are the signs of the 1QC and 2QC spectra (see Fig.~\ref{fig:spectra}).  

  In order to compare the magnitudes of the~$\kappa$QC peaks let us introduce
  the shorthand
  \begin{equation}\label{eq:def-peak-amp}
    \PeakAmp{k}{\parallel/\perp}{\kappa}
    =
    \PeakAmpExplicit{k}{\parallel/\perp}{\kappa}
  \end{equation}
  for the amplitude of the real part of these spectra. Note that the
  dependence on the pulse area is not spelled out explicitly for brevity.

For a single fine transition (D2-line) of rubidium atoms, the positions,
lineshapes, and signs of all resonances in Fig.~\ref{fig:spectra}, as well as
the absence of the 1QC signals in the $\perp$ channel are in qualitative agreement 
with the experimental observations\cite{Bruder_phd,Bruder_2019}.
Furthermore, the ratio of the peak values of 2QC and 1QC signals,
$
  \PeakAmp{x}{\parallel}{2}
  /
  \PeakAmp{x}{\parallel}{1}
  \sim
  \num{e-2}
$, 
approximately equals the experimental value\cite{Bruder_2019}. 
Finally, the calculated inhomogeneously broadened spectral lines have absorptive parts with dominantly Gaussian line shapes, whose full widths at half maximum are about $100\gamma$, or $\approx \SI{1.39}{GHz}$. This also agrees, in
order of magnitude, with the experimentally observed line widths\cite{Bruder_phd}.

Let us briefly recap which are the crucial ingredients to  capture the essential physics and to
obtain reasonable qualitative and quantitative agreement with experiment: 
\begin{itemize}
\item First of all, to obtain the $\kappa$QC spectra for different channels and
observation directions, one needs to incorporate the vector nature of atomic
dipole transitions, of which the here considered atoms equipped with a
$\JG=0\leftrightarrow \JE=1$ transition represent the simplest example. 
\item Second, performing the configuration average is extremely important to account for
both, the randomness of atomic positions, and their drift in a thermal
gas.
Without this procedure, neither the line shapes nor their magnitudes have any
resemblance with the observed ones. 
For instance, for individual random realizations, both $S_{\uvec{x}}(\omega;1)$
and $S_{\uvec{y}}(\omega;1)$ would yield non-zero contributions in the $\perp$
channel, and it is only the configuration average which completely suppresses this
specific signal. 
\item Third, given the diluteness of the atomic sample, the dominant contribution to
the dipole-dipole interaction is due to its far-field part describing real 
photon exchange.
By considering only the near-field (electrostatic) term instead, one
substantially underestimates the magnitude of MQC signals.
More importantly, unlike the electrostatic interaction which brings about
collective levels shifts, the far-field dipole-dipole interactions lead to
both, collective shifts \emph{and} collective decay processes.
Let us note in passing that collective decay plays a crucial role even for very
close emitters, where it can trigger coherent excitation flow under incoherent
driving\cite{shatokhin2018_njp}.
\end{itemize}

\subsection{Dependence of the peak amplitudes on the pulse area}
\label{sec:area}
Let us now exploit the potential of our non-perturbative approach and examine
the behavior of the real parts of the 1QC and 2QC spectra for $N=2$ atoms under changes of the
pulse area~$\vartheta$. 

\bgroup
\def\tt{\frac{\vartheta}{2}}
\def\t{\vartheta}
\def\xx{\bar{\xi}^2}
\def\sig{\bar{\Delta}}
\begin{table}
  \renewcommand{\arraystretch}{1.5}
  \begin{tabular}{%
      >{$}c<{$}
      >{$}c<{$}
      !{\hspace{-.3em}}
      >{$}c<{$}
      !{\hspace{.1em}}
      |
      !{\hspace{.2em}}
      >{$}l<{$}
    }
    \toprule
      \kappa &
      \uvec{k} &
      \text{\parbox{2em}{chan\-nel}} &
      \PeakAmp{k}{\parallel\!/\!\perp}{\kappa}
      \big/
      \bigl(f^2 / \sqrt{2\pi} \gamma^2\bigr)
      \\
    \midrule
    1 & \uvec{x} & \parallel &
    \begin{aligned}
      &
      \textstyle
      \frac{1}{80 \xx}
      \sin^2\t
      \Bigl[
	\frac{3 \gamma^2}{\sig^2} \cos^3\tt
	\\
      &
      \textstyle
      \hspace{.7em} {} - 3 V\!\bigl(\frac{\gamma}{2\sig}\bigr) \cos\tt \Bigl(
	\!
	    \frac{\gamma^2}{\sig^2} + (1 - 4\cos\tt - \cos\t)
	    \!
	  \Bigr)
      \\
      &
      \textstyle
      \hspace{.7em} {}+ V\!\bigl(\frac{3 \gamma}{2\sig}\bigr) \sin^2\tt \Bigl(
	\!
	  \frac{3 \gamma^2}{\sig^2} \cos\tt + (2\cos\tt - 4\cos\t)
	  \!
	  \Bigr)
	  \!
	\Bigr]
      \end{aligned}
	\\
     1 & \uvec{y} & \parallel &
	V\!({\gamma}/{\sqrt{2} \sig})
	\sin^2\t
	+ \mathscr{O}\bigl(\bar{\xi}^{-2}\bigr)
      \\
     1 &
     \uvec{x}, \uvec{y} &
     \perp &
     0
    \\
    \midrule
     2 & \uvec{x} & \parallel &
	-\frac{3}{320 \xx}
	V\!({\gamma}/{\sqrt{2} \sig})
	\sin^4\t
	\\
     2 & \uvec{y} & \parallel &
	-\frac{51}{640 \xx}
	V\!({\gamma}/{\sqrt{2} \sig})
	\sin^4\t
	\\
     2 & \uvec{x}, \uvec{y} & \perp &
	-\frac{3}{320 \xx}
	V\!({\gamma}/{\sqrt{2} \sig})
	\sin^2\tt \sin^2\t
	\\[.5ex]
    \bottomrule
  \end{tabular}
  \caption{%
    \label{tab:analytical-peaks}
    Peak amplitudes
    $
      \PeakAmp{k}{\parallel/\perp}{\kappa}
      =
      \PeakAmpExplicit{k}{\parallel/\perp}{\kappa}
    $
    of the single \eqref{S1} and double quantum coherence spectra \eqref{S2} of $N=2$ fluorescing atoms,
    with
    $
      V\!(x)
      =
      \sqrt{\pi/2} \, x \, \exp(x^2/2) \erfc(x/\sqrt{2})
    $,
    the r.m.s.\ Doppler shift
    $\sig = k_\rmL \sqrt{k_\mathrm{B} T / M}$
    defined through~\eqref{MBdistr}
    and the constant~$f$ given below~\eqref{Eplus}.
    For each signal we give here the leading order, which is always $\mathscr{O}\bigl(\bar{\xi}^{-2}\bigr)$ except for $\PeakAmp{y}{\parallel}{1}$, where the single-scattering contribution dominates as $\mathscr{O}(0)$, see Fig.~\ref{fig:scattering} (a).
}
\end{table}
\egroup

\begin{figure}
  \includegraphics[width=0.5\textwidth]{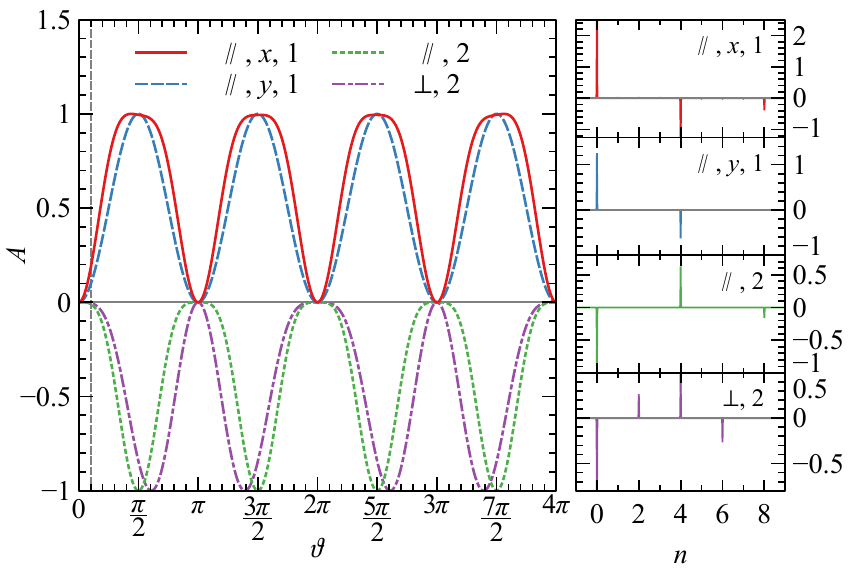}
  \caption{
     \label{fig:peaks}
    (Left) Peak amplitudes
    $
      \PeakAmp{k}{\parallel/\perp}{\kappa}
      =
      \PeakAmpExplicit{k}{\parallel/\perp}{\kappa}
    $
    of the single and double quantum coherence spectra of $N=2$ fluorescing atoms (same color code as Fig.~\ref{fig:spectra}),
    obtained from table~\ref{tab:analytical-peaks} by 
    normalizing their amplitude to $\pm 1$
      (hence, the 2QC signals' amplitudes
      $
	\PeakAmp{x}{\parallel}{2}
	\propto
	\PeakAmp{y}{\parallel}{2}
      $
      and
      $
	\PeakAmp{x}{\perp}{2}
	=
	\PeakAmp{y}{\perp}{2}
      $
      are represented by one green dotted and one purple dash-dotted line, respectively),
    as functions of the pulse area $\vartheta$, for
    $\parallel$ and $\perp$ pump-probe polarizations and detection directions
    $\uvec{k}=\uvec{x},\uvec{y}$ (see legends).
    The vertical dashed line at $\vartheta=\num{0.14}\pi$ corresponds to the pulse area
    in Fig.~\ref{fig:spectra}.
    The values of the other parameters are the same as in Fig.~\ref{fig:spectra}.
    (Right) Amplitudes $A_n$ of the trigonometric
    expansion~\eqref{expand} of~$A(\vartheta)$ as displayed in the left panel
    (identified by the color code). 
  }
\end{figure}

  In table~\ref{tab:analytical-peaks}, we provide the expressions for the
  peak amplitudes~$\PeakAmp{k}{\parallel\!/\!\perp}{\kappa}$ as defined
  in~\eqref{eq:def-peak-amp}, to leading order in~$\bar{\xi}^{-2}$.
These expressions are oscillatory functions of~$\vartheta$
with periods varying from $\pi$ (for
$\PeakAmp{y}{\parallel}{1}$,
$\PeakAmp{x}{\parallel}{2}$, and
$\PeakAmp{y}{\parallel}{2}$),
over $2\pi$ (for
$\PeakAmp{x}{\perp}{2}$ and
$\PeakAmp{x}{\perp}{2}$),
to $4\pi $ (for
$\PeakAmp{x}{\parallel}{1}$).
Due to their different scaling with the inverse square of the scaled
interatomic distance, $\bar{\xi}^{-2}$, and/or different numerical prefactors,
these functions vary significantly in magnitude (see also the variable scales
of the $y$-axis in Fig.~\ref{fig:spectra}).
Therefore, to compare their behavior solely as functions of $\vartheta$,
Fig.~\ref{fig:peaks} (left) normalizes the functions to unit amplitude, retaining the overall sign.

We notice that all functions vanish at $\vartheta=n\pi$ ($n=0,1,2,\ldots$).
The occurrence of the zeroes at integer multiples of~$\pi$ 
can be understood as an immediate consequence of the Rabi dynamics of coherently driven two-level atoms \cite{allen_eberly}, together with
the fact that the atoms are interacting with
two identical time-delayed pulses.
Since typically these delays $\tau\sim \SI{10}{\pico\second}$ are short (see
Sec.~\ref{sec:timescales}),
spontaneous emission events between the pulses can be ignored.
Thus, the first $\pi$-pulse \cite{allen_eberly} coherently transfers the atoms from the ground to
the excited state, while the second $\pi$-pulse, conversely, coherently brings
the atoms back to their ground states.
Hence, the atoms cannot fluoresce, regardless of
MQC order $\kappa$, polarization channel, or observation direction. 

According to the same logic, the largest peak magnitudes can be expected at odd
integer multiples of the ``optimal'' pulse area,
$\vartheta_\mathrm{opt}=\pi/2$.
After two such pulses, the atomic population of an isolated atom is coherently
transferred  from the ground to the excited state, feeding a fluorescence signal which
attains a maximum.
And indeed, we observe maximal magnitudes for the functions
$\PeakAmp{y}{\parallel}{1}$,
$\PeakAmp{x}{\parallel}{2}$ and
$\PeakAmp{y}{\parallel}{2}$
at
$\vartheta=(2n+1)\vartheta_\mathrm{opt}$
($n=0,1,2,\ldots$), see Fig.~\ref{fig:peaks}~(left).
However, the functions
$\PeakAmp{x}{\parallel}{1}$ and
$
  \PeakAmp{x}{\perp}{2}
  =
  \PeakAmp{y}{\perp}{2}
$
reach their maximal magnitudes for the first time at values of~$\vartheta\approx 0.4\pi$ and $\approx 0.6\pi$, respectively, that are slightly
shifted with respect to the ``optimal'' pulse area $0.5\pi$. 
We attribute this to peculiar physical mechanisms for the generation of 1QC or
2QC signals, which require excited state degeneracy and collective decay
processes\cite{ames2021}. 

In order to further quantify the distinctions between the peak amplitudes, in
Fig.~\ref{fig:peaks}~(right) we have plotted the coefficients $A_n$ obtained by
expansion of the oscillatory functions from Fig.~\ref{fig:peaks}~(left) in a
trigonometric series
\begin{equation}
  A(\vartheta)
  =
  \sum_{n}A_n\cos\frac{n\vartheta}{2}. 
  \label{expand}
\end{equation}
Despite quite similar periodic behavior of the peak amplitudes Fig.~\ref{fig:peaks}~(left), their expansions \eqref{expand} elucidate the differences between distinct signals via the coefficients $A_n$. In other words, 
these coefficients can be conceived as specific ``fingerprints''
reflecting the order~$\kappa$, the
polarizations of the laser pulses, as well as the observation direction. Thus, the nonlinear laser-atom interaction processes induced by strong driving fields modify the MQC signals, and this modification is in itself a diagnostic tool of dipolar interactions in dilute thermal gases.

\section{Conclusions}
\label{sec:conclusions}
We expanded our non-perturbative open-system theory of multiple quantum coherence\cite{ames20,beni_MSc} to account for thermal atomic motion, and to address the dependence of MQC signals on the driving strength as quantified by the incoming pulses' areas. While the former leads to improved qualitative agreement between our numerical results for $N=2$ atoms (as the minimal model to incorporate all physically relevant processes which contribute to the detected fluorescence) and experiment\cite{lukas_bruder15}, the latter defines a new diagnostic tool to discriminate different excitation channels, beyond the perturbative regime.

\acknowledgements
E.~G.~C.~acknowledges the support of the G.~H.~Endress Foundation.
V.~S.~and A.~B.~thank the Strategiefonds der Albert-Ludwigs-Universit\"at
Freiburg for partial funding.

\section*{DATA AVAILABILITY}
The data that supports the findings of this study are available within the article.

\bibliography{dip_dip_hooray}

\end{document}